\documentclass[11pt,a4paper]{article}
\usepackage[margin=2.5cm]{geometry}
\usepackage{amsmath,amssymb}
\usepackage{braket}
\usepackage[colorlinks=true,linkcolor=blue,citecolor=blue,urlcolor=blue]{hyperref}

\newcommand{\SL}{\mathrm{SL}(2,\mathbb{C})}
\newcommand{\Y}{Y_\gamma}
\newcommand{\dd}{\mathrm{d}}
\newcommand{\Tr}{\operatorname{Tr}}
\newcommand{\pair}[2]{\langle #1,\, #2\rangle}

\title{What the flatness constraint of EPRL really constrains}
\author{Claudio Perini\thanks{Center for SpaceTime and the Quantum (CSTQ), Cassis, France}}
\date{}

\begin{document}

\maketitle

\begin{abstract}
On a fixed 2-complex the EPRL spin foam model seems to allow only flat geometries, or
accidentally curved ones with deficit angles $4\pi/\gamma$ apart. We derive the
classical action exactly and find that the model is BF theory with two constraints, one imposed
by a delta on the labels and the other by an exponential on the gluing. The delta turns the two flatness
equations of a face into one, on a combination of its curvature and its twist;
the exponential makes that one equation sharp, and fixes the curvature only to
the width of the gluing. The Barbero--Immirzi parameter sets that width. The
small-$\gamma$ regime of the literature is this width at work, and it comes with
a scale: lowering $\gamma$ allows the same curved geometry on a coarser triangulation.
\end{abstract}

\section{Introduction}
\label{sec:intro}

The flatness problem of the EPRL spin foam model is the following observation. On a fixed
2-complex, when the spins are large and summed over, the stationary points of
the amplitude are flat geometries \cite{Bonzom,HellmannKaminski},
or geometries with accidental curvature \cite{HellmannKaminski}, whose deficit
angles are multiples of $4\pi/\gamma$ \cite{Han5628}. The curved Regge
geometries that the vertex amplitude reproduces one simplex at a time
\cite{BDFHP} seem to be lost when the simplices are glued. Two readings have
been given: that the model needs a modification, or that the large-spin limit
at fixed complex is not where the classical theory should be looked for, and
that curvature returns when $\gamma$ is small \cite{MagliaroPerini,Han5628}.

This note takes the second reading and makes it quantitative, by writing the
model as a path integral over a classical action. The spin sum can be done
exactly, and the result is BF theory with two constraints: one imposed sharply
by a delta on the representation labels, the other weakly by an exponential on
the gluing of the normals. Everything the model does with curvature follows
from these two. Section~\ref{sec:action} derives the action and sets it next to
BF theory. Section~\ref{sec:circle} explains the integers in it with a particle
on a circle. Section~\ref{sec:gluing} follows the gluing exponential to the
widths it produces, and Section~\ref{sec:gamma} collects what the
Barbero--Immirzi parameter does with them. The derivations are in the appendix.

\section{The classical action of EPRL}
\label{sec:action}

We start from the EPRL model itself \cite{EPRL}. It is defined on a 2-complex dual to a
triangulation, whose vertices, edges and faces are dual to 4-simplices, tetrahedra and
triangles. With one spin $j_f$ per face,
one $\SL$ element $g_{ve}$ per half-edge, and the wedge transports
$G_k=g_{ve_{k+1}}^{-1}g_{ve_k}$ going around each face, the amplitude is
\begin{equation}
Z=\sum_{\{j_f\}}\prod_fd_{j_f}\int\prod\dd g_{ve}\;\prod_f
\Tr\Big[\prod_{k\in f}\Y^\dagger D^{(\gamma j_f,\,j_f)}(G_k)\Y\Big],
\qquad d_j=2j+1 ,
\label{eq:eprl}
\end{equation}
where $\Y$ embeds the spin-$j$ representation of $\mathrm{SU}(2)$ into the
principal series representation $(\gamma j,j)$ of $\SL$. Lattice $\SL$ BF
theory, $Z_{\rm BF}=\int\prod\dd g\prod_f\delta(U_f)$, has the same form:
expanding the delta function on the principal series gives \eqref{eq:eprl} with
the full representation $(\rho,\kappa)$ in place of $\Y^\dagger(\cdot)\Y$, both
labels summed with the Plancherel weight. Write the representations on
homogeneous functions of a spinor \cite{Ruhl}. Then the trace of a face becomes
an integral over one unit vector $\hat u_k$ per wedge, and the two face
integrands are, wedge by wedge,
\begin{subequations}
\label{eq:compare}
\begin{align}
\text{BF:}\qquad&\prod_k\int\dd\mu(\hat u_k)\;\;
\underbrace{\delta_{S^2}\big(\hat u_{k+1},\hat u_k'\big)}_{\text{gluing, sharp}}
\;\;\underbrace{e^{\,i(\rho\,\eta_k+2\kappa\,\xi_k)}}_{\text{two labels }\rho,\,\kappa}\;\;e^{-\eta_k},
\label{eq:compare-bf}\\[6pt]
\text{EPRL:}\qquad&\prod_k\int\dd\mu(\hat u_k)\;\;
\underbrace{(2j+1)\,e^{-2j\lambda_k}}_{\text{gluing, weak}}
\;\;\underbrace{e^{\,i\,j(\gamma\,\eta_k+2\,\xi_k)}}_{\text{one label, }\rho=\gamma \kappa}\;\;e^{-\eta_k}.
\label{eq:compare-eprl}
\end{align}
\end{subequations}
Here $\eta_k=2\ln\|G_ku_k\|$ is the boost across the wedge, $\xi_k$ is the
rotation of the frame of the triangle about its own normal, which we call the twist,
and $\lambda_k\ge0$ measures by how much the transported normal $\hat u_k'$ misses the
next one, $\hat u_{k+1}$: the factor $e^{-2j\lambda_k}=|\braket{u_{k+1}|u_k'}|^{2j}$
is a Gaussian of width $1/\sqrt j$ around $\hat u_{k+1}=\hat u_k'$. Nothing is
approximated, and the derivation is in Appendix~\ref{app:trace}. The exponent of
the EPRL integrand is exactly linear in $j$, because the homogeneity degree is linear in $j$.

Same normals, same $\eta_k$ and $\xi_k$, same measure, same form of the phase.
The two theories differ in the two marked places and nowhere else. In the labels: BF has two
independent ones, $\rho$ paired with the boost and $\kappa$ paired with the twist, and
EPRL has one, $\rho=\gamma\kappa$; this is a delta on the labels, the diagonal
simplicity constraint, and it is what turns the two angles into the one composite
angle. In the gluing: BF has the delta function $\delta_{S^2}(\hat u_{k+1},\hat u_k')$,
which forces the transported normal to be the next one, and EPRL has the
exponential $(2j+1)e^{-2j\lambda_k}$ in its place, the same condition imposed
weakly, with width $1/\sqrt j$. The exponential is the reproducing kernel of the
projector $\Y\Y^\dagger$ onto $V_j$, so this is the transverse simplicity
constraint, and it is the whole origin of the weight. EPRL is BF theory with a
delta on the labels and an exponential on the gluing, and nothing else. As far
as we know, this exact relation to BF theory is not in the literature.

The sum over the compact label can now be done, in both theories, by Poisson
resummation (Appendix~\ref{app:poisson}): the label becomes a continuous variable plus an integer $m_f$ per
face. For EPRL the label is the spin and the continuous variable is the area
$A_f=\gamma j_f$; for BF it is $\kappa_f$, while $\rho_f$ is continuous already. With
$\mathcal H_f$, $\Xi_f$, $\Lambda_f$ the sums of $\eta_k$, $\xi_k$, $\lambda_k$
around the face, and with the integrals over normals and transports, the factor
$e^{-\sum_k\eta_k}$ and the polynomial measures left implicit, the two path integrals are
\begin{subequations}
\label{eq:Scl}
\begin{align}
\text{BF:}\qquad&\sum_{\{m_f\}}\int\dd\rho\,\dd\kappa
\underbrace{\prod_k\delta_{S^2}\big(\hat u_{k+1},\hat u_k'\big)}_{\text{gluing, sharp}}
\exp i\sum_f\Big[\underbrace{\rho_f\,\mathcal H_f+\kappa_f\,\big(2\Xi_f+4\pi m_f\big)}_{\text{two multipliers}}\Big],
\label{eq:Scl-bf}\\[6pt]
\text{EPRL:}\qquad&\sum_{\{m_f\}}\int\dd A
\underbrace{e^{-2\sum_f(A_f/\gamma)\Lambda_f}}_{\text{gluing, weak}}
\exp i\sum_f\Big[\underbrace{\tfrac{A_f}{\gamma}\,\big(\gamma\mathcal H_f+2\Xi_f+4\pi m_f\big)}_{\text{one multiplier}}\Big].
\label{eq:Scl-eprl}
\end{align}
\end{subequations}
The same two
differences appear, and now they have a consequence. The labels multiply
the angles linearly, so they are Lagrange multipliers, and the two theories differ
in their equations of motion. In BF, varying $\rho_f$ and $\kappa_f$ gives two equations
per face, $\mathcal H_f=0$ and $2\Xi_f+4\pi m_f=0$. There is no boost around the
face and the rotation about the normal is trivial, so the face is flat. In EPRL there is one
multiplier, the area, and one equation. The action is
\begin{equation}
S_{\rm cl}=\frac1\gamma\sum_fA_f\,\vartheta_f,\qquad
\vartheta_f=\gamma\mathcal H_f+2\Xi_f+4\pi m_f,
\label{eq:action}
\end{equation}
and varying the area gives
\begin{equation}
\frac{\partial S_{\rm cl}}{\partial A_f}=0\;\Longleftrightarrow\;\vartheta_f=0 ,
\label{eq:eom}
\end{equation}
the flatness constraint. We call $\vartheta_f$ the composite angle of the face. So the model is a path
integral over areas, normals and transports, with a real action that is linear in
the areas, times a positive weight that is largest when the normals glue; and the
one equation it imposes per face, a trivial composite angle, is the sum of the two
flatness equations of BF, not either of them. Notice that both lines of
\eqref{eq:Scl} are first-order area--angle actions \cite{DittrichSpeziale,Bonzom}: in BF each face has two areas, one seen by the boost part of its
bivector and one by the rotation part, each conjugate to its own angle; in EPRL
the delta on the labels makes them one area, conjugate to one composite angle.

This also says what the action is. Lattice BF theory pairs the flux of each face
with the logarithm of the holonomy, $S_{\rm BF}=\sum_f\pair{J_f}{\log U_f}$, and
varying the six components of the flux gives six flatness equations per face. Take
the flux to be simple before varying, $J_f=(A_f\hat u,\,(A_f/\gamma)\hat u)$, which
is the EPRL condition $\vec K=\gamma\vec L$. Then the pairing keeps one angle out of
six, the boost along the normal plus $\gamma$ times the rotation about it, with the
weights of the Holst flux $\star B+B/\gamma$. This is $\vartheta_f$, and it is the
class angle of the Ashtekar--Barbero holonomy around the hinge (Appendix~\ref{app:class}). On a Regge geometry
the twist vanishes and the pairing is $\pm A_f\Theta_f$, with
$\Theta_f$ the deficit angle of the face, which is its curvature. This is the Regge
action, at every spin and with no limit taken.

\section{The winding sectors: a particle on a circle}
\label{sec:circle}

One may object at this point: an action with an integer summed over is not a
classical action: something quantum has been left in it. It has not. The integers
label different classical histories, and the sum over them is the sum over
histories that every path integral contains. The cleanest way to see it is a
particle on a circle. Take one face of \eqref{eq:Scl-eprl}, with the weight omitted, and
put it next to the propagator of a free particle on a circle, written in the same
way, as a sum over windings and an integral over the momentum:
\begin{subequations}
\label{eq:circle}
\begin{align}
\text{particle:}\qquad K&=\sum_{n\in\mathbb Z}\int\dd p\;
\exp i\Big[\;\underbrace{p}_{\text{momentum}}\;\big(\underbrace{\Delta\theta}_{\text{angle}}
+\underbrace{2\pi n}_{\text{winding}}\big)\;-\!\!\underbrace{p^2T/2}_{\text{Hamiltonian}}\Big],
\label{eq:circle-particle}\\[8pt]
\text{face:}\qquad Z_f&=\sum_{m\in\mathbb Z}\int\dd A\;
\exp i\Big[\;\underbrace{\tfrac{A}{\gamma}}_{\text{momentum}}\;\big(\underbrace{\gamma\mathcal H+2\Xi}_{\text{angle}}
+\underbrace{4\pi m}_{\text{winding}}\big)\;-\!\!\underbrace{\;0\;}_{\text{Hamiltonian}}\Big].
\label{eq:circle-face}
\end{align}
\end{subequations}
The area over $\gamma$ is the momentum, the composite angle is the angle swept, the
integer is the winding, and there is no Hamiltonian. Two things can be read off
this display.

The sectors are kinematical. The sum over $n$ is there before any equation of
motion is written: it is the sum over the paths that wind $n$ times, and the
propagator contains all of them. The sum over $m_f$ is the same thing. The integers
are the branches of the logarithm of the holonomy in its compact direction; they are
in \eqref{eq:Scl} whatever the constraints; they are different histories and not
copies of one another, and the sum over them must be kept.

The sectors get their curvature on shell. For the particle, stationarity in $p$
gives each sector its own momentum; for the face, stationarity in $A$ does not fix
the area but fixes the angle, and solving the angle of sector $m$ for the boost
gives each sector its own deficit:
\begin{equation}
\text{particle:}\quad p_n=\frac{\Delta\theta+2\pi n}{T},
\qquad\qquad
\text{face:}\quad \mathcal H_f=-\,\frac{2\Xi_f+4\pi m_f}{\gamma}.
\label{eq:sectors}
\end{equation}
For the particle, different windings are different classical paths with different
momenta, so with different energies, and no one calls the sum over them a quantum
leftover: it is the sum over classical histories. For the face it is the same
sentence with one word changed: different integers are different classical
histories with different boost deficits, spaced by $4\pi/\gamma$. The accidental
curvatures of the flatness literature \cite{Bonzom,HellmannKaminski,Han5628} are the on-shell values of the winding
sectors, as the $p_n$ are the on-shell momenta of the windings of the particle, and
the sum over $m_f$ is the sum over curved histories, not a trace of the spins.

\section{The flatness constraint and the gluing}
\label{sec:gluing}

Everything the model imposes weakly, it imposes through one factor, the
exponential on the gluing. Section~\ref{sec:action} showed that this factor is the only
difference between EPRL and BF theory besides the tie of the labels. So if the
flatness constraint has a width, and if the curvature has a width, both must come
from there. This section follows that factor to its consequences.

Start with the flatness constraint. For one face, the integral over the area runs
against the phase and the weight together,
\begin{equation}
\int\dd A_f\;\exp\Big[\frac{A_f}{\gamma}\big(i\,\vartheta_f-2\Lambda_f\big)\Big],
\label{eq:faceint}
\end{equation}
with the sum over the windings understood. Where the normals glue exactly,
$\Lambda_f=0$, this is a delta in $\vartheta_f$: the composite angle is set to
zero exactly, as the uniform sum over the momentum of a particle with no
Hamiltonian is a delta in the angle. Where they do not, it is a peak of width
$2\Lambda_f$ (Appendix~\ref{app:faceint}).

How wide is the gluing? Call a crossing the edge--face pair $(e,f)$ where the face
passes from one wedge to the next, and let $\phi_k$ be its mismatch angle, the
angle by which the transported normal misses the next one. The weight of one crossing is a Gaussian in
that angle,
\begin{equation}
e^{-2(A/\gamma)\lambda_k},\qquad
\lambda_k=-\ln\cos\frac{\phi_k}{2}\simeq\frac{\phi_k^2}{8},
\qquad\text{of width}\qquad
\phi\;\sim\;\sqrt{\gamma/A}\,.
\label{eq:phi}
\end{equation}
We call $\sqrt{\gamma/A}$ the gluing width. The mismatch $\Lambda_f$ is a sum of
terms $\phi_k^2/8$, so the width of the flatness constraint \eqref{eq:eom} is
the square of the gluing width. This is what we mean by a sharp constraint and
a weak gluing.

A mismatch is produced by any change of the transport $G_k$ that tilts the
normal: a boost along a direction $\hat m$ transverse to it, a rotation about
$\hat m$, or both. The weight allows mismatch angles up to the gluing width.

Now start from a Regge geometry and make such a change at one tetrahedron $e$:
apply to its half-edge element
$g_{ve}$ a boost of rapidity $\epsilon$ along $\hat m$ and a rotation by an angle
$\alpha$ about $\hat m$, with $\hat m$ transverse to the normal of the face $f$
whose gluing is being loosened. The element $g_{ve}$ enters the wedge
transports of all four faces through $e$, so the change reaches each of them,
but only in the one wedge that contains it. The three angles of a face, its boost $\mathcal H$, its twist $\Xi$ and their
combination $\vartheta$, are sums around its loop, so each of them changes by the
change of that one wedge. For
$f$ itself nothing changes at first order, the change being transverse to its
own normal. For another face $f'$ through $e$, with $\chi$ the 3d dihedral angle
between the two triangles, the response of its angles is (Appendix~\ref{app:response})
\begin{equation}
\delta\mathcal H=\epsilon\sin\chi,\qquad
\delta(2\Xi)=\alpha\sin\chi,\qquad
\delta\vartheta=(\gamma\epsilon+\alpha)\sin\chi,
\label{eq:invariance}
\end{equation}
all at first order: the boost moves the curvature, the rotation moves the twist,
and the composite angle moves with both. So the gluing lets the curvature and
the twist of a face each move by $\sqrt{\gamma/A}$. The flatness constraint
\eqref{eq:eom} is stricter:
it holds $\vartheta$ within $\Lambda_f$, which is quadratic in the mismatch,
$\gamma/A$, the square of the gluing width. Of all the moves the gluing allows,
the constraint keeps only the simplicity transformation, a boost with a rotation
by $\gamma$ times the rapidity about the same axis, $\alpha=-\gamma\epsilon$.
Under it the curvature moves and the twist follows, but the composite angle does
not move.

The width of the angles and that of the constraint both follow at once from the gluing width,
\begin{equation}
\underbrace{\delta\mathcal H,\;\delta(2\Xi)\;\sim\;\phi\;\sim\;\sqrt{\gamma/A}}_{\text{angles: first order}},
\qquad\qquad
\underbrace{\delta\vartheta\;\sim\;\Lambda\;\sim\;\phi^2\;\sim\;\gamma/A}_{\text{constraint: second order}},
\label{eq:widths}
\end{equation}
so that $\delta\vartheta\sim(\delta\mathcal H)^2$: at large area the width of the
constraint is much smaller than the response of the angles. So the composite
angle is fixed and the curvature is not: the constraint $2\Xi=-\gamma\mathcal H$ ties the
torsion of the face, that is its twist \cite{Wieland}, to the curvature sharply, and
fixes neither. So the
flatness constraint fixes the composite angle, not the curvature; the curvature
is fixed only as far as the gluing is, and the gluing is weak. How much
curvature the gluing lets through is computed in the next section.

At infinite area only the critical configurations survive \cite{BDFHP}, where the normals
glue exactly: the constraint is a delta, the composite angle is zero, and,
since the twist vanishes, so does the curvature. This gives flatness, or accidental
curvature. At finite area the deficit angles acquire a width \cite{Han5628}; its
origin is the gluing.

The same could be said in canonical language, which we do not develop here. The
simplicity transformation would be a gauge direction, generated by the part of
simplicity along the normal, and the composite angle would be invariant under it.
The transverse part of simplicity would be second class, which is why the model
can impose it only weakly \cite{EPRL}.

\section{The Barbero--Immirzi parameter at work}
\label{sec:gamma}

The parameter $\gamma$ enters the model in one place, the composite angle
$\vartheta=\gamma\mathcal H+2\Xi$, and does one thing there: it converts a boost
into a rotation. Everything else it does follows from this conversion.

Take the simplicity transformation: a boost by $\epsilon$ along a direction
transverse to the normal, together with a rotation by $\gamma\epsilon$ about it. In the composite
angle the two cancel: a boost by $\epsilon$ counts as a twist by $\gamma\epsilon$.
In the flatness constraint, $2\Xi=-\gamma\mathcal H$, $\gamma$ trades torsion for
curvature. It is not the multiplier of the constraint, which is the area, but
its coupling: it decides how much curvature the constraint sees.

The gluing width, found in the previous section, is $\sqrt{\gamma/A}$, and
$\gamma$ sets it: at fixed area, a smaller $\gamma$ glues more tightly, a larger
$\gamma$ more loosely. The shapes of
the tetrahedra match within this width \cite{BDFHP}. A Regge geometry with deficit angle
$\Theta$ has $\vartheta=\gamma\Theta$. By \eqref{eq:invariance} a twist shifts
$\vartheta$ by the same amount, and the gluing allows twists up to
$\sqrt{\gamma/A}$. So the geometry sits within the width of the constraint as
long as $\gamma\Theta\lesssim\sqrt{\gamma/A}$ \cite{Han5628,ADH} (Appendix~\ref{app:deficit}):
\begin{equation}
\delta\phi\;\sim\;\sqrt{\gamma/A}\,,\qquad\qquad
\delta\Theta\;\sim\;\frac{\delta\phi}{\gamma}\;\sim\;\frac{1}{\sqrt{\gamma A}}\,.
\label{eq:width}
\end{equation}
As $\gamma$ grows the first width grows and the second shrinks: $\gamma$ decides
which width is the larger, that of the shapes or that of the deficit angles. At
small $\gamma$ the shapes match and the deficit angles of a Regge geometry are
free. This is the small-$\gamma$ regime of \cite{MagliaroPerini,Han5628}, in
which the model is Regge calculus with unconstrained deficit angles, and curved
geometries appear among the critical points \cite{HanHuangLiuQu}. At large
$\gamma$ the shapes are loose and the deficit angles are held to a lattice,
which is the flatness problem. This is the dynamical role of
$\gamma$, and it lives in the exponent of the amplitude, in the imaginary part
that curved configurations acquire.

The Barbero--Immirzi parameter also enters something that is purely kinematical, the compactness of
the twist. The twist is an angle, with
period $4\pi$; the boost is not periodic. In the composite angle the two are added
with the conversion factor, so the period of the twist becomes, for the deficit
angle, a period $4\pi/\gamma$, the window of curvature a sector can hold: the
accidental curvatures are the branches of a
compact torsion angle read as curvature, and this is why their spacing is
$1/\gamma$. The same compactness makes the areas discrete, with gap $\gamma/2$,
and gap times window is $2\pi$, a fixed quantum that $\gamma$ apportions between
the fineness of the area spectrum and the range of curvature one sector can hold.
A deficit angle beyond the window is not suppressed: like a frequency beyond the
Nyquist limit, it is read as a smaller one in another sector. The same cutoff
is known for the Ashtekar--Barbero holonomy \cite{CharlesLivine}.
This is the kinematical role of $\gamma$.

We end with the interplay between $\gamma$ and refinement. Take a large, smooth, curved geometry with curvature
radius $R$, and triangulate it. A face of area $A$ carries a deficit angle
$\Theta\sim A/R^2$. Now coarsen the triangulation: the faces grow, and the deficit
on each of them grows with them. Two things happen on the way. First, the deficit
angles can exceed their width: the model admits them within
$\delta\Theta\sim1/\sqrt{\gamma A}$, which shrinks as the faces grow while the deficit
grows, so beyond $\gamma A^3\sim R^4$
the deficit angles are pinned to the lattice $4\pi\mathbb Z/\gamma$ and the
geometry is read as flat or accidentally curved. Then, at $\gamma A\sim4\pi R^2$,
the deficit reaches the window itself and is aliased: the curvature of the smooth
geometry is counted as a different, smaller curvature in another sector. The
large-area limit at fixed complex walks exactly this road, since it scales the
areas with the complex held fixed: it is the regime in which the model does not
see the smooth geometry, not the regime in which it describes it. Refine instead:
the deficit per face shrinks faster than its width, the deficit angles are free,
and the model describes the curved geometry, until the faces reach the quantum
of area, $A\sim\gamma$, where the gluing width $\sqrt{\gamma/A}$ is of order one
and the shapes dissolve. So a smooth geometry is described faithfully only on a range of
face sizes,
\begin{equation}
\gamma\;\ll\;A\;\ll\;\gamma^{-1/3}R^{4/3}\,:
\label{eq:range}
\end{equation}
areas large enough for the shapes to match, small enough for the deficit
angles to be free.

The flatness problem is the statement that the range \eqref{eq:range} has an upper end, where
the triangulation is too coarse; the model's answer is that geometry lives inside
the range, and that a triangulation that is too coarse can only be rescued by
lowering $\gamma$. Lower $\gamma$ to $\gamma/N$, keeping the geometry fixed. The deficit angles
are free as long as $\gamma A^3\ll R^4$, so the same geometry can now be
triangulated with faces $N^{1/3}$ times larger without losing its curvature. The lower end drops as well, and this is the deep quantum side: the area gap
shrinks from $\gamma/2$ to $\gamma/2N$, and a face of given size holds $N$ times more quanta. It
is no surprise that the two go together. The limit in which the areas become
continuous is the limit in which the deficit angles become free on coarse
triangulations. It is the limit in which the model decompactifies to Regge
calculus, where the areas are continuous and the deficit angle is no longer read
through a compact one.

\section{Outlook}
\label{sec:outlook}

The widths are given here only up to factors of order one; their coefficients
remain to be computed and compared with the semiclassical analyses and with
effective spin foam models. The refinement argument of the previous section is
about regimes. It says where a smooth geometry sits relative to the widths of
the faces, not what a sum over the bulk of a refined complex does. Whether
an actual coarse graining of the model is compatible with it is not known. A
canonical treatment of the face, in which the weak gluing appears as the weak
imposition of second-class constraints, would make the canonical reading of
Section~\ref{sec:gluing} precise.

\section*{Acknowledgments}

We thank Carlo Rovelli for reading the manuscript and for useful discussions.

\appendix

\section{Derivations}
\label{app}

\subsection{The face trace}
\label{app:trace}

We realise the principal series $(\rho,\kappa)$ of $\SL$ on functions of a spinor
$z\in\mathbb C^2$, $z\ne0$, homogeneous of degree
$(n_1,n_2)=(-1+i\rho+\kappa,\,-1+i\rho-\kappa)$ \cite{Ruhl}:
\begin{equation}
f(az)=a^{n_1}\bar a^{\,n_2}f(z),\qquad (D(G)f)(z)=f(G^Tz).
\end{equation}
The inner product is an integral over the sphere of directions; on unit spinors
we write its measure $\dd\mu(\hat u)$, normalised to one. An operator $A$ acts
through a kernel, $(Af)(z)=\int\dd\mu(\hat w)\,K_A(z,w)f(w)$. The kernel of
$D(G)A$ is $K_A(G^Tz,w)$, the kernel of a product is the composition of the
kernels, and the trace is the integral of the kernel on the diagonal. For a face,
\begin{equation}
\Tr\Big[\prod_kD(G_k)P\Big]=\int\prod_k\dd\mu(\hat z_k)\;\prod_kK_P\big(G_k^Tz_k,z_{k+1}\big),
\qquad z_{n+1}\equiv z_1 ,
\label{eq:app-chain}
\end{equation}
where $P$ is the operator at each edge of the face: the identity for BF theory, the
projector $\Y\Y^\dagger$ for EPRL. On unit spinors the two kernels are
\begin{equation}
K_1(z,w)=\delta_{S^2}(\hat z,\hat w)\,e^{2i\kappa\arg\braket{w|z}},\qquad
K_Y(z,w)=(2j+1)\braket{w|z}^{2j},
\label{eq:app-KY}
\end{equation}
with $\braket{w|z}=\bar w_0z_0+\bar w_1z_1$, and off unit spinors they follow
from homogeneity. The first is the kernel of the identity; its phase follows from the homogeneity
of degree $(n_1,n_2)$. The
second, with $(\rho,\kappa)=(\gamma j,j)$, reproduces the spin-$j$ block, the
holomorphic polynomials of degree $2j$, as follows from
$\int\dd\mu\,|w_0|^{2a}|w_1|^{2b}=a!\,b!/(a+b+1)!$.

Now write $G_k^Tz_k=\|G_k^Tz_k\|\,\hat z_k'$, with $\eta_k=2\ln\|G_k^Tz_k\|$ and
$e^{-\lambda_k+i\xi_k}=\braket{z_{k+1}|\hat z_k'}$. Homogeneity takes the norm out
as $e^{(-1+i\rho)\eta_k}$, and
\begin{subequations}
\label{eq:app-kernels}
\begin{align}
K_1(G_k^Tz_k,z_{k+1})&=e^{(-1+i\rho)\eta_k}\,\delta_{S^2}(\hat z_k',\hat z_{k+1})\,e^{2i\kappa\xi_k},
\label{eq:app-K1}\\
K_Y(G_k^Tz_k,z_{k+1})&=(2j+1)\,e^{(-1+i\gamma j)\eta_k}\,e^{-2j\lambda_k}\,e^{2ij\xi_k}.
\label{eq:app-KYG}
\end{align}
\end{subequations}
These are \eqref{eq:compare-bf} and \eqref{eq:compare-eprl}. The phase of each $z_k$ cancels between the two kernels that
contain it, so only the directions $\hat u_k$ remain; in the main text $G_ku_k$
stands for $G_k^Tz_k$. For BF theory the face traces are summed over both labels
with the Plancherel formula,
$\delta(U)=\sum_\kappa\int\dd\rho\,(\rho^2+\kappa^2)\,\chi_{(\rho,\kappa)}(U)$ up
to a constant.

\subsection{The resummation}
\label{app:poisson}

The spins are half-integers. For a function $g$ on the half-line, Poisson
summation on $\mathbb Z/2$ gives
\begin{equation}
\sum_{j\in\mathbb N/2}g(j)=\tfrac12 g(0)+\sum_{m\in\mathbb Z}\,2\!\int_0^\infty\!\dd j\;g(j)\,e^{4\pi imj} ,
\end{equation}
where the term $\tfrac12g(0)$ comes from cutting the lattice at $j=0$. With
$A=\gamma j$ the phase $e^{4\pi imj}=e^{i(A/\gamma)4\pi m}$ adds $4\pi m_f$ to the
composite angle, which is the integer of \eqref{eq:action}. We drop the term
$\tfrac12g(0)$, the contribution of zero area. In BF the label $\kappa$ runs over
all half-integers, and the same formula holds without that term.

The period $4\pi$ comes from the twist. For half-integer $j$, $e^{2ij\Xi}$ does
not change when $\Xi$ grows by $2\pi$, that is when $2\Xi$ grows by $4\pi$, and
the integer counts these turns. The period refers to a fixed choice of sign of
the half-edge elements. Replacing $g_{ve}$ by $-g_{ve}$ does not change the
amplitude, because the spins of a tetrahedron add up to an integer, but it
shifts by $2\pi$ the composite angles of the four faces through the
tetrahedron. The spacing $2\pi/\gamma$ of \cite{HellmannKaminski} counts both
signs; the spacing $4\pi/\gamma$, as in \cite{Han5628}, holds them fixed.

\subsection{The composite angle as a class angle}
\label{app:class}

Where the normals glue, $\hat u_{k+1}=\hat u_k'$, the transport of a wedge acts
on the spinor of the normal as $G_ku_k=e^{\eta_k/2+i\xi_k}u_{k+1}$. Around the
face, $U_fu_1=e^{\mathcal H_f/2+i\Xi_f}u_1$. So $u_1$ is an eigenvector of the
face holonomy with eigenvalue $e^{(\mathcal H_f+2i\Xi_f)/2}$, and the holonomy
is conjugate to a boost by $\mathcal H_f$ composed with a rotation by $2\Xi_f$
about the same axis. The Ashtekar--Barbero connection $\Gamma+\gamma K$ replaces
the boost generator by $\gamma$ times the rotation generator. In this class the
result is a rotation by $\gamma\mathcal H_f+2\Xi_f$, which is $\vartheta_f$ up to
the branch $4\pi m_f$. On a Regge geometry the frame of each triangle is
transported without rotation about its normal, so $\Xi_f=0$ and $\mathcal H_f$
is the deficit angle.

\subsection{The face integral}
\label{app:faceint}

At fixed transports and normals, $\vartheta_f$ and $\Lambda_f\ge0$ are numbers,
and the area integral of \eqref{eq:faceint} is
\begin{equation}
\int_0^\infty\dd A\;e^{(A/\gamma)(i\vartheta-2\Lambda)}=\frac{\gamma}{2\Lambda-i\vartheta} .
\label{eq:app-lorentz}
\end{equation}
Its real part, $2\gamma\Lambda/(4\Lambda^2+\vartheta^2)$, is a Lorentzian in
$\vartheta$ centred at zero, of half-width $2\Lambda$. As $\Lambda\to0$ the
integral becomes $\gamma[\pi\delta(\vartheta)+i\,\mathrm{P}(1/\vartheta)]$, a
delta in $\vartheta$ plus a principal value. With the discrete areas kept,
$A=\gamma j$, the sum over the spins is
\begin{equation}
\sum_{j\in\mathbb N/2}e^{j(i\vartheta-2\Lambda)}=\frac{1}{1-e^{(i\vartheta-2\Lambda)/2}}
=\frac12+\sum_{m\in\mathbb Z}\frac{2}{2\Lambda-i(\vartheta+4\pi m)}\,.
\end{equation}
This is the same peak repeated with period $4\pi$, one copy for each winding,
plus the endpoint term of Appendix~\ref{app:poisson}. The factors $2j+1$ of the
measure add derivatives with respect to $\vartheta$ and do not move the peaks.

\subsection{The response of the angles}
\label{app:response}

Take a Regge geometry and a tetrahedron $e$ at a vertex $v$. On a Regge geometry
each wedge transport carries the two opposite directions $\pm\hat u$ of a normal
to the two opposite directions $\pm\hat u'$ at the next tetrahedron. So it maps
the spinor $u$ of the normal to a multiple of $u'$, and the orthogonal spinor
$u^\perp$ to a multiple of $u'^\perp$.

Let $\hat u$ be the normal of the face $f$ in the frame of $e$, let
$\hat m\perp\hat u$, and change the half-edge element by
$g_{ve}\to g_{ve}\,e^{X}$, with $X=(\epsilon+i\alpha)\,\hat m\cdot\vec\sigma/2$:
a boost of rapidity $\epsilon$ along $\hat m$ and a rotation by $\alpha$ about
$\hat m$. In the frame of $e$ this multiplies by $e^X$ the spinor of every face
of $e$ before it is transported.

For a face $f'$ of $e$ with normal $\hat u'$, at first order
$e^Xu'=u'+\tfrac12(\epsilon+i\alpha)(\hat m\cdot\vec\sigma)u'$. The spinor
$(\hat m\cdot\vec\sigma)u'$ has the component $(\hat m\cdot\hat u')\,u'$ along
$u'$ and a component along $u'^\perp$. The first multiplies $u'$ by
$1+\tfrac12(\epsilon+i\alpha)\,\hat m\cdot\hat u'$: its modulus changes the boost,
$\delta\eta=\epsilon\,\hat m\cdot\hat u'$, and its phase changes the twist,
$\delta(2\xi)=\alpha\,\hat m\cdot\hat u'$. The second tilts the direction of
$u'$, which is a mismatch; since the transport maps $u'^\perp$ to a multiple of
the orthogonal spinor at the next tetrahedron, it does not change $\eta$ or
$\xi$ at first order. With $\hat m$ in the plane of $\hat u$ and $\hat u'$, one
has $\hat m\cdot\hat u'=\sin\chi$, with $\chi$ the dihedral angle between the two
triangles, and this is \eqref{eq:invariance}.

For the face $f$ itself $\hat m\cdot\hat u=0$, and only the tilt remains. Its
angle is $\sqrt{\epsilon^2+\alpha^2}$ at $e$, and
$e^{-\eta}\sqrt{\epsilon^2+\alpha^2}$ at the next tetrahedron, where it is
compared with the normal $\hat u_{k+1}$ of that tetrahedron; $\eta$ is the boost of the wedge. Summing
$(A_{f'}/\gamma)\,\delta\vartheta_{f'}$ over the four faces of $e$, the
first-order change of the phase is
$(\gamma\epsilon+\alpha)\,\hat m\cdot\sum_{f'}A_{f'}\hat u_{f'}/\gamma$, for any
direction $\hat m$. Stationarity under all such changes is the closure of the
tetrahedron.

\subsection{The width of the deficit angle}
\label{app:deficit}

Take a Regge geometry with a deficit angle $\Theta$ on the face $f'$, so that
$\vartheta_{f'}=\gamma\Theta$. By \eqref{eq:invariance}, a rotation by $\alpha$
at a crossing of $f'$ shifts $\vartheta_{f'}$ by $\alpha\sin\chi$ and leaves the
curvature unchanged. It also produces mismatch angles of order $\alpha$ on the
faces of the tetrahedron. The configuration with $\alpha\sin\chi=-\gamma\Theta$
satisfies the flatness constraint, and its weight is
$\exp[-c\,\gamma A\,\Theta^2]$, with $c$ a geometric factor of order one when the
dihedral angles and the boosts are. This is a Gaussian in $\Theta$ of width of
order $1/\sqrt{\gamma A}$, which is \eqref{eq:width}. A boost at the crossing
also shifts $\vartheta_{f'}$, but it shifts $\mathcal H_{f'}$ with it, and the
configuration that satisfies the flatness constraint is then flat. The curved
configurations that contribute are the twisted ones.

\end{document}